\documentclass[12pt,prd,aps,amssymb,amsmath,tightenlines,showpacs]{revtex4}
\usepackage{psfig}

\usepackage{graphics}
\usepackage[pdftex]{graphicx}
\newcommand{\half}{\mbox{$\textstyle\frac{1}{2}$}}
\begin{document}
\preprint{}

\title{Semiclassical Calculation of the $\mathcal{C}$ Operator in
$\mathcal{PT}$-Symmetric Quantum Mechanics}

\author{Carl~M.~Bender\footnote{Permanent address: Department of Physics,
Washington University, St. Louis, MO 63130, USA.} and Hugh~F.~Jones}

\affiliation{Blackett Laboratory, Imperial College, London SW7 2BZ, UK}

\date{\today}

\begin{abstract}
To determine the Hilbert space and inner product for a quantum theory defined by
a non-Hermitian $\mathcal{PT}$-symmetric Hamiltonian $H$, it is necessary to
construct a new time-independent observable operator called $C$. It has recently
been shown that for the {\it cubic} $\mathcal{PT}$-symmetric Hamiltonian $H=p^2+
x^2+i\epsilon x^3$ one can obtain $\mathcal{C}$ as a perturbation expansion in
powers of $\epsilon$. This paper considers the more difficult case of noncubic
Hamiltonians of the form $H=p^2+x^2(ix)^\delta$ ($\delta\geq0$). For these
Hamiltonians it is shown how to calculate $\mathcal{C}$ by using nonperturbative
semiclassical methods.
\end{abstract}


\maketitle

\section{Introduction}
\label{s1}

In 1998 it was discovered that the class of non-Hermitian Hamiltonians
\begin{eqnarray}
H=p^2+x^2(ix)^\delta\qquad(\delta>0)
\label{e1}
\end{eqnarray}
has a positive real spectrum~\cite{BB}. It was conjectured in Ref.~\cite{BB}
that the spectral positivity was associated with the space-time reflection
symmetry ($\mathcal{PT}$ symmetry) of the Hamiltonian. The Hamiltonian
(\ref{e1}) is $\mathcal{PT}$ symmetric because $x\to-x$ and $p\to-p$ under
parity reflection $\mathcal{P}$, and $x\to x$, $p\to-p$, and $i\to-i$ under time
reversal $\mathcal{T}$. Many other $\mathcal{PT}$-symmetric quantum mechanical
models have been examined~\cite{BBM,R1,R2,R3,R4}, and a proof of the positivity
of the spectrum of $H$ in (\ref{e1}) was subsequently given by Dorey {\it et
al.} \cite{DDT}.

Once the positivity and reality of the spectrum of a $\mathcal{PT}$-symmetric
Hamiltonian $H$, such as the non-Hermitian Hamiltonian (\ref{e1}), has been
established one must then demonstrate that $H$ defines a consistent unitary
theory of quantum mechanics. To do so, one shows that the Hilbert space on which
the Hamiltonian acts has an inner product associated with a positive norm and
that the time evolution induced by such a Hamiltonian is unitary; that is, the
norm is preserved in time \cite{AM,BBJ}. Specifically, for a complex Hamiltonian
having an {\it unbroken} $\mathcal{PT}$ symmetry, one must construct a linear
operator $\mathcal{C}$ that commutes with both $H$ and $\mathcal{PT}$. One can
then show that the inner product with respect to $\mathcal{CPT}$ conjugation
satisfies the requirements for the theory defined by $H$ to have a Hilbert space
with a positive norm and to be a consistent unitary theory of quantum mechanics.
[The term {\it unbroken} ${\cal PT}$ symmetry means that every eigenfunction of
$H$ is also an eigenfunction of the ${\cal PT}$ operator. This condition
guarantees that the eigenvalues of $H$ are real. The Hamiltonian in (\ref{e1})
has an unbroken $\mathcal{PT}$ symmetry for all real $\delta\geq0$.]

In a conventional quantum theory the inner product is formulated with respect to
ordinary Dirac Hermitian conjugation (complex conjugate and transpose). Unlike
the situation with conventional quantum theory, the inner product for a quantum
theory defined by a non-Hermitian $\mathcal{PT}$-symmetric Hamiltonian depends
on the Hamiltonian itself and is thus determined dynamically. One must first
find the eigenstates of $H$ before knowing what the Hilbert space and the
associated inner product of the theory are. The Hilbert space and inner product
are then determined by these eigenstates.

We emphasize that the key breakthrough in understanding these non-Hermitian
$\mathcal{PT}$-symmetric quantum theories was the discovery of the operator
$\mathcal{C}$ \cite{BBJ}. The problem is therefore to construct $\mathcal{C}$
for a given $H$. In Refs.~\cite{BBJ,BBJ2} it was shown how to express the
$\mathcal{C}$ operator in coordinate space as a formal sum over appropriately
normalized eigenfunctions $\phi_n(x)$ of the Hamiltonian $H$. These
eigenfunctions satisfy the Schr\"odinger equation
\begin{eqnarray}
H\phi_n(x)=E_n\phi_n(x),
\label{e2}
\end{eqnarray}
and, without loss of generality, their overall phases are chosen so that
\begin{eqnarray}
\mathcal{PT}\phi_n(x)=\phi_n(x).
\label{e3}
\end{eqnarray}
With this choice of phase, the eigenfunctions are then normalized according to
\begin{eqnarray}
\int_C dx\,[\phi_n(x)]^2=(-1)^n.
\label{e4}
\end{eqnarray}
The contour of integration $C$ is described in detail in Ref.~\cite{BBJ}. For
the quantum theories described by the Hamiltonian (\ref{e1}), $C$ can be taken
to lie along the real-$x$ axis if $\delta<2$.

In terms of the eigenfunctions defined above, the statement of completeness for
a theory described by a non-Hermitian $\mathcal{PT}$-symmetric Hamiltonian is
\cite{BBJ}
\begin{eqnarray}
\sum_n(-1)^n\phi_n(x)\phi_n(y)=\delta(x-y)
\label{e5}
\end{eqnarray}
for real $x$ and $y$. The unusual factor of $(-1)^n$ arises because of the
convention (\ref{e3}). The coordinate-space representation of $\mathcal{C}$
is~\cite{BBJ}
\begin{eqnarray}
\mathcal{C}(x,y)=\sum_n\phi_n(x)\phi_n(y).
\label{e6}
\end{eqnarray}
Only a {\it non-Hermitian} $\mathcal{PT}$-symmetric Hamiltonian possesses a
$\mathcal{C}$ operator distinct from the parity operator $\mathcal{P}$. Indeed,
if one evaluates the summation (\ref{e6}) for a $\mathcal{PT}$-symmetric
Hamiltonian that is also Hermitian, the result is $\mathcal{P}$, which in
coordinate space is $\delta(x+y)$.

The coordinate-space formalism using (\ref{e6}) has been applied successfully to
the cubic Hamiltonian
\begin{eqnarray}
H=\half p^2+\half\mu^2 x^2+i\epsilon x^3,
\label{e7}
\end{eqnarray}
and $\mathcal{C}$ was constructed perturbatively to order $\epsilon^3$
\cite{BMW}. The approach in Ref.~\cite{BMW} was to calculate the eigenfunctions
as perturbation series in powers of $\epsilon$ and then to substitute these
series into (\ref{e6}). The sum was then evaluated order by order in $\epsilon$.
This technique was also used to calculate $\mathcal{C}$ to order $\epsilon$ for
the cubic complex H\'enon-Heiles Hamiltonian \cite{HH,BBRR}
\begin{eqnarray}
H=\half\left(p_x^2+p_y^2\right)+\half\left(x^2+y^2\right)+i\epsilon x^2y,
\label{e8}
\end{eqnarray}
which has two degrees of freedom, and for the Hamiltonian
\begin{eqnarray}
H=\half\left(p_x^2+p_y^2+p_z^2\right)+\half\left(x^2+y^2+z^2\right)+i\epsilon
xyz,
\label{e9}
\end{eqnarray}
which has three degrees of freedom \cite{BBRR}.

Calculating the operator $\mathcal{C}$ by evaluating the sum in (\ref{e6})
directly is difficult in quantum mechanics and hopeless in quantum field theory
because to do this it is necessary to determine all the eigenfunctions of $H$.
In Refs.~\cite{AAAA,BBBB} we showed that there is a simpler way to calculate
$\mathcal{C}$ based on three crucial properties of this operator. First,
$\mathcal{C}$ is $\mathcal{PT}$-symmetric (that is, it commutes with the 
space-time reflection operator $\mathcal{PT}$),
\begin{eqnarray}
[\mathcal{C},\mathcal{PT}]=0,
\label{e10}
\end{eqnarray}
although $\mathcal{C}$ does not commute with $\mathcal{P}$ or $\mathcal{T}$
separately. Second, the square of $\mathcal{C}$ is the identity,
\begin{eqnarray}
\mathcal{C}^2={\bf 1},
\label{e11}
\end{eqnarray}
which allows us to interpret $\mathcal{C}$ as a reflection operator. Third,
$\mathcal{C}$ commutes with $H$,
\begin{eqnarray}
[\mathcal{C},H]=0,
\label{e12}
\end{eqnarray}
and thus is time independent.

We also observed that there is a natural way to represent $\mathcal{C}$ as an
exponential of a real Hermitian operator $Q$ multiplying the parity operator
$\mathcal{P}$~\cite{AAAA,BBBB}:
\begin{eqnarray}
\mathcal{C}=e^{Q(x,p)}\mathcal{P},
\label{e13}
\end{eqnarray}
where $x$ and $p$ are the dynamical operator variables. (This exponential
representation was first noticed in Ref.~\cite{BMW}.) The advantage of this
representation is that when $\mathcal{C}$ is written in this form, properties
(\ref{e10}) and (\ref{e11}) are equivalent to the symmetry conditions that $Q(x,
p)$ be an {\it even} function of $x$ and an {\it odd} function of $p$. The 
remaining condition (\ref{e12}) yields an operator equation that determines
$Q(x,p)$. This operator equation can be solved perturbatively for Hamiltonians
having a cubic interaction term. The perturbative procedure is so easy that the
$\mathcal{C}$ operator can even be found for cubic quantum field theories such
as \cite{AAAA,BBBB}
$$H=\half\pi^2+\half(\nabla\varphi)^2+\half\mu^2\varphi^2+i\epsilon\varphi^3.$$

To illustrate the perturbative procedure, we show how to find $\mathcal{C}$ for
the Hamiltonian (\ref{e7}). We expand the operator $Q(x,p)$ as a series in odd
powers of $\epsilon$ \cite{AAAA,BBBB}:
\begin{eqnarray}
Q(x,p)=\epsilon Q_1(x,p)+\epsilon^3 Q_3(x,p)+\epsilon^5Q_5(x,p)+\ldots\,.
\label{e14}
\end{eqnarray}
We then substitute (\ref{e14}) into (\ref{e12}) and collect the coefficients of
like powers of $\epsilon^n$ for $n=1,2,3,\ldots$. The result is a sequence of
equations that can be solved systematically for the operator-valued functions
$Q_n(x,p)$ $(n=1,3,5,\ldots)$ subject to the symmetry constraints that ensure
the conditions (\ref{e10}) and (\ref{e11}). The first three
of these equations read
\begin{eqnarray}
\left[H_0,Q_1\right] &=& -2H_1,\nonumber\\
\left[H_0,Q_3\right] &=& -{\textstyle\frac{1}{6}}[Q_1,[Q_1,H_1]],\nonumber\\
\left[H_0,Q_5\right] &=& {\textstyle\frac{1}{360}}[Q_1,[Q_1,[Q_1,[Q_1,H_1]]]]
-{\textstyle\frac{1}{6}}\left([Q_1,[Q_3,H_1]]+[Q_3,[Q_1,H_1]]\right),
\label{e15}
\end{eqnarray}
where $H_0=\frac{1}{2}p^2+\frac{1}{2}\mu^2x^2$ and $H_1=ix^3$. We now substitute
the most general polynomial form for $Q_n$ using arbitrary coefficients and then
solve for these coefficients. For example, to solve the first of the equations
in (\ref{e15}), we take as an {\it ansatz} for $Q_1$ the most general real
Hermitian cubic polynomial that is even in $x$ and odd in $p$:
\begin{eqnarray}
Q_1(x,p)=Mp^3+Nxpx,
\label{e16}
\end{eqnarray}
where $M$ and $N$ are undetermined coefficients. The operator equation for $Q_1$
is satisfied if
\begin{eqnarray}
M=-{\textstyle\frac{4}{3}}\mu^{-4}\quad{\rm and}\quad N=-2\mu^{-2}.
\label{e17}
\end{eqnarray}
In Ref.~\cite{BBJ} we determined the $\mathcal{C}$ operator to seventh order in
perturbation theory. We were able to perform this high-order calculation because
our procedure does not make use of the summation in (\ref{e6}), and therefore we
did not need to find the wave functions $\phi_n(x)$.

Unfortunately, this perturbative procedure for finding the operator $\mathcal{C
}$ is ineffective when the Hamiltonian is not cubic because this approach leads
to complicated and unwieldy infinite sums \cite{AAAA}. Furthermore, the
perturbative procedure does not work in the massless limit. [Note that the
coefficients in (\ref{e17}) diverge when $\mu=0$.] Thus, a more powerful method
is needed to find the $\mathcal{C}$ operator associated with the Hamiltonian
(\ref{e1}).

In this paper we introduce a completely new approach based on nonperturbative
semiclassical methods rather than perturbative methods. The procedure is
straightforward; the WKB physical-optics approximation to the eigenfunctions
$\phi_n(x)$ is substituted into the formula (\ref{e6}) for the operator
$\mathcal{C}$ and the summation is performed. This procedure is justified
because the WKB approximation is asymptotically accurate in the limit of large
$n$. We will see that when the WKB eigenfunctions are used to evaluate this sum,
the result is a singular operator. The error incurred by including the small-$n$
terms in the sum is finite, and it therefore may be neglected.

In the next section we show how to perform the WKB calculation of the
$\mathcal{C}$ operator and in Sec.~\ref{s3} we discuss the properties of our
solution for $\mathcal{C}$ and consider the possible extension of this work to
quantum field theory.

\section{WKB Calculation of the $\mathcal{C}$ Operator}
\label{s2}

We begin this section with a review of the WKB analysis of the conventional 
two-turning-point problem. We then generalize the standard treatment of the
two-turning-point problem from the real-$x$ axis to the complex plane, where we
apply it to the Hamiltonian (\ref{e1}).

\subsection{WKB on the Real Axis}
The conventional treatment of the two-turning-point eigenvalue problem on the
real axis is described in Ref.~\cite{BO}. This problem is expressed in terms of
the differential equation
\begin{eqnarray}
-\epsilon^2 \phi''(x)+V(x)\phi(x)=E\phi(x),
\label{e18}
\end{eqnarray}
where $\epsilon\ll1$ is a small positive perturbation parameter. The parameter
$\epsilon$ is used to organize the WKB expansion and to distinguish between
different orders in the semiclassical perturbation theory. We assume that the
potential $V(x)$ rises as $x\to\pm\infty$ and that there are two turning points
at $x=A$ and at $x=B$ with $A<B$. The turning points $A$ and $B$ are solutions
to the algebraic equation $V(x)=E$. The classically allowed region is $A<x<B$
and the classically forbidden regions are $x<A$ and $x>B$. The quantization
condition on the eigenvalue $E$ in (\ref{e18}) is the requirement that the wave
function $\phi(x)$ vanish as $x\to\pm\infty$. This quantization condition leads
to a discrete spectrum $E_n$ ($n=0,\,1,\,2,\,\cdots$).

For fixed $n$ and small $\epsilon$, WKB theory gives a good approximation to the
eigenvalues $E_n$ and the corresponding eigenfunctions $\phi_n(x)$. (For fixed
$\epsilon$ the WKB approximation becomes accurate as $n\to\infty$.) To leading
order in WKB theory (the {\it physical-optics} approximation) the quantization
condition for the eigenvalues $E_n$ reads
\begin{eqnarray}
\frac{1}{\epsilon}\int_{A_n}^{B_n}dx\,\sqrt{E_n-V(x)}\sim\left(n+\frac{1}{2}
\right)\pi\quad(\epsilon\to0^+).
\label{e19}
\end{eqnarray}
Also, the physical-optics approximation to the wave function $\phi_n(x)$ in the
classically allowed region $A_n<x<B_n$ is given by
\begin{eqnarray}
\phi_n(x)\sim C\left[E_n-V(x)\right]^{-1/4}\sin\left[\frac{1}{\epsilon}
\int_{A_n}^x dt\,\sqrt{E_n-V(t)}+\frac{\pi}{4}\right]\quad(\epsilon\to0^+).
\label{e20}
\end{eqnarray}
The complete WKB physical-optics approximation to the wave function $\phi_n(x)$
actually consists of five separate formulas, but to calculate the operator
$\mathcal{C}$ we will not need the approximations to the wave function in the
regions near the turning points at $A_n$ and $B_n$ and in the two classically
forbidden regions. (These formulas are given in Ref.~\cite{BO}.)

The multiplicative constant $C$ in (\ref{e20}) is determined by the usual
normalization condition that the integral of the square of the eigenfunction be
unity:
\begin{eqnarray}
1=\int_{-\infty}^\infty dx\,\phi_n^2(x)\sim\frac{1}{2}C^2\int_A^B dx\,\left[E_n
-V(x)\right]^{-1/2}\quad(\epsilon\to0^+).
\label{e21}
\end{eqnarray}
The calculation that leads to this asymptotic evaluation of the integral is
explained in detail in Ref.~\cite{BO}. Note that the condition in (\ref{e21})
employs the conventional normalization criterion that is used for Hermitian
Hamiltonians rather than the normalization used in (\ref{e4}).

\subsection{WKB in the Complex Plane}

The Schr\"odinger eigenvalue equation (\ref{e2}) associated with the Hamiltonian
(\ref{e1}) is 
\begin{eqnarray}
-\epsilon^2\phi_n''(x)+x^2(ix)^\delta\phi_n(x)=E_n\phi_n(x).
\label{e22}
\end{eqnarray}
We have inserted the small positive parameter $\epsilon$ to organize the
structure of the semiclassical approximation but at the end of the calculation
we will set $\epsilon=1$ and use the fact that WKB becomes accurate as $n\to
\infty$ to justify our results.

\begin{figure}[b!]
\vspace{3.0in}
\includegraphics{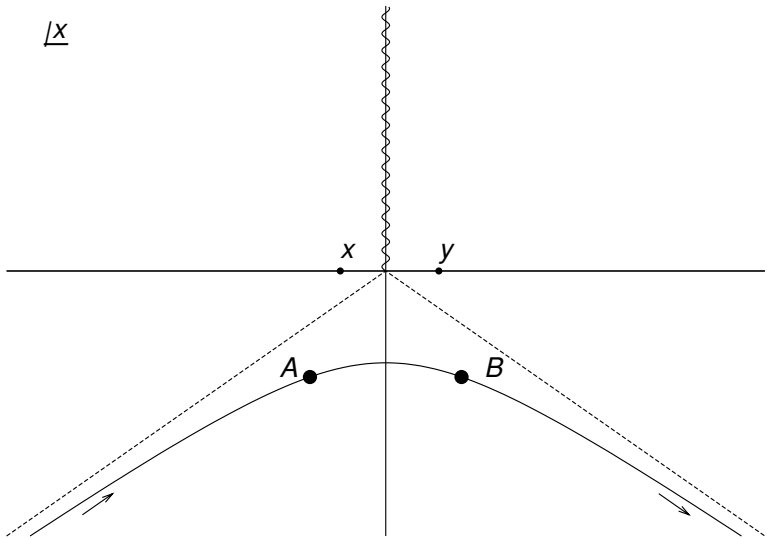}
\caption{Schematic representation of the contour (solid line) in the complex-$x$
plane along which the WKB approximation to the Schr\"odinger equation
(\ref{e22}) is derived. This contour passes through the two turning points $A$
and $B$ given in (\ref{e23}). These turning points are indicated by dots. For
large $|x|$ the WKB contour approaches the centers of the asymptotic wedges
(dashed lines) in which the eigenfunctions vanish exponentially as $|x|\to
\infty$. The eigenfunctions are oscillatory on the WKB contour between $A$ and
$B$. (This region is the complex generalization of the classically allowed
region in conventional WKB theory.) Outside of this oscillatory region the
eigenfunctions decay exponentially along the WKB contour. (These regions are the
complex generalizations of the classically forbidden regions.) The
eigenfunctions are analytic in the complex-$x$ plane except on the branch cut,
which runs along the imaginary-$x$ axis from the origin to $i\infty$. To
calculate the $\mathcal{C}$ operator the WKB approximation (\ref{e20}) to the
eigenfunction $\phi_n$ is evaluated on the real-$x$ axis at the points $x$ and
$y$, which are at a distance of order $1$ from the origin.}
\label{f1}
\end{figure}

To construct a physical-optics approximation to the eigenfunctions $\phi_n(x)$
for this equation, we begin by finding the turning points $A_n$ and $B_n$, which
satisfy $E_n=x^2(ix)^\delta$:
\begin{eqnarray}
A_n=E_n^\frac{1}{2+\delta}\,e^{-i\pi+\frac{i\pi\delta}{4+2\delta}}\quad{\rm and}
\quad B_n=E_n^\frac{1}{2+\delta}\,e^{-\frac{i\pi\delta}{4+2\delta}}.
\label{e23}
\end{eqnarray}
These turning points lie on the real axis when $\delta=0$ and rotate downward
into the complex-$x$ plane as $\delta$ increases from 0. (The turning points
are shown in Fig.~\ref{f1}.)

The WKB analysis is done along a contour passing through these turning points.
This contour, which is shown as a solid line in Fig.~\ref{f1}, lies in the
lower-half complex-$x$ plane in the asymptotic Stokes wedges in which the
eigenfunctions vanish exponentially as $|x|\to\infty$. For large $|x|$ the WKB
contour is asymptotic to the centers of the wedges; these centers are shown on
Fig.~\ref{f1} as dashed lines. In the quadrant ${\rm Re}\,x>0$, ${\rm Im}\,x<0$
the center of the wedge lies at the angle $-\frac{\delta}{8+2\delta}\pi$; the
upper and lower edges of the wedge (not shown in Fig.~\ref{f1}) lie at the
angles $-\frac{\delta-2}{8+2\delta}\pi$ and $-\frac{\delta+2}{8+2\delta}\pi$. A
detailed description of the WKB contour and the asymptotic wedges is given in
Ref.~\cite{BB}.

The WKB formulas (\ref{e19}) -- (\ref{e21}) for the eigenfunctions $\phi_n(x)$
are valid in the complex-$x$ plane as well as on the real-$x$ axis. Our
ultimate objective is to use these formulas to evaluate the sum (\ref{e6}) and
thereby to determine the $\mathcal{C}$ operator. However, to demonstrate the
techniques needed to evaluate this sum we will first show how to evaluate the
sum in (\ref{e5}) that expresses the completeness condition and we will verify
the (approximate) completeness of the WKB wave functions in (\ref{e20}).

We begin our analysis by finding the WKB approximation to the product $\phi_n(x)
\phi_n(y)$. We will then perform the sum in (\ref{e5}) under the assumption that
the arguments $x$ and $y$ of the eigenfunctions are real and of order 1. We must
take $x$ and $y$ to be real because, as we will see, they will appear as
arguments of a Dirac delta function, and the delta function is only defined for
real argument. (The points $x$ and $y$ are shown in Fig.~\ref{f1}.) We will
assume that $n\gg1$, and therefore that $E_n\gg1$. Under this assumption we can
make the approximations
\begin{eqnarray}
\left[E_n-x^2(ix)^\delta\right]^{-1/4}\sim E_n^{-1/4}\quad(\epsilon\to0^+)
\label{e24}
\end{eqnarray}
and
\begin{eqnarray}
\int_{A_n}^x dt\,\sqrt{E_n-t^2(it)^\delta}\sim I_n+\sqrt{E_n}x\quad(\epsilon
\to0^+),
\label{e25}
\end{eqnarray}
where
\begin{eqnarray}
I_n=\int_{A_n}^0 dt\,\sqrt{E_n-t^2(it)^\delta}=\frac{\sqrt{\pi}}{2}E_n^\frac{4
+\delta}{4+2\delta}\frac{\Gamma\left(\frac{3+\delta}{2+\delta}\right)}{\Gamma
\left(\frac{8+3\delta}{4+2\delta}\right)}e^\frac{i\pi\delta}{4+2\delta}.
\label{e26}
\end{eqnarray}
Using the trigonometric identity $2\sin\left(a+\frac{\pi}{4}\right)\sin\left(b+
\frac{\pi}{4}\right)=\cos(a-b)+\sin(a+b)$, we obtain the WKB approximation to
the product $\phi_n(x)\phi_n(y)$:
\begin{eqnarray}
\phi_n(x)\phi_n(y)&\sim&\frac{(-1)^n}{\sqrt{E_n}\int_{A_n}^{B_n}dt\,\left[E_n-
t^2(ix)^\delta\right]^{-1/2}}\left\{\cos\left[\frac{1}{\epsilon}\sqrt{E_n}(x-y)
\right]\right.\nonumber\\
&&\qquad\left.+\sin\left[\frac{2}{\epsilon}I_n+\frac{1}{\epsilon}\sqrt{E_n}(x+y)
\right]\right\}\quad(\epsilon\to0^+),
\label{e27}
\end{eqnarray}
where the factor of $(-1)^n$ comes from the normalization condition (\ref{e4}).

We will see that this factor of $(-1)^n$ plays a crucial role in our evaluation
of sums because oscillatory terms are suppressed relative to nonoscillatory
terms. This factor arises because the phase-fixing condition (\ref{e3}) requires
that the eigenfunctions $\phi_n(x)$ be $\mathcal{PT}$ symmetric. The harmonic
oscillator ($\delta=0$) illustrates clearly the origin of the $(-1)^n$ factor.
Requiring that the harmonic oscillator eigenfunctions be $\mathcal{PT}$
symmetric (that is, that they be functions of $ix$) means that they have the
general form $\phi_n(x)=i^ne^{-x^2/2}He_n(x)$, where $He_n(x)$ is the usual
Hermite polynomial. It is the extra factor of $i^n$ that gives rise to the
factor of $(-1)^n$ in the normalization condition (\ref{e4}). The factor
$(-1)^n$ persists for all $\delta>0$ \cite{BBJ}.

Next, we use the quantization condition (\ref{e19}) to obtain the identity
\begin{eqnarray}
\frac{2}{\epsilon}I_n=\left(n+\frac{1}{2}\right)\pi+\frac{1}{\epsilon}(I_n
-I_n^*),
\label{e28}
\end{eqnarray}
where $*$ denotes complex conjugation. This identity allows us to rewrite
(\ref{e27}) as
\begin{eqnarray}
\phi_n(x)\phi_n(y)&\sim&\frac{1}{\sqrt{E_n}\int_{A_n}^{B_n}dt\,\left[E_n-t^2
(ix)^\delta\right]^{-1/2}}\left\{(-1)^n\cos\left[\frac{1}{\epsilon}\sqrt{E_n}
(x-y)\right]\right.\nonumber\\
&&\qquad\left.+\cos\left[\frac{1}{\epsilon}(I_n-I_n^*)+\frac{1}{\epsilon}
\sqrt{E_n}(x+y)\right]\right\}\quad(\epsilon\to0^+),
\label{e29}
\end{eqnarray}

The next step is to perform the sum over $n$. We do so by converting the sum to
an integral using the general WKB density of states result
\begin{eqnarray}
dE_n\,\frac{1}{2\epsilon}\int_{A_n}^{B_n} dx\,\left[E_n-V(x)\right]^{-1/2}
=\pi\,dn\quad(\epsilon\to0^+),
\label{e30}
\end{eqnarray}
which is obtained by differentiating the quantization condition (\ref{e19}).
Note that one must differentiate the endpoints as well as the integrand with
respect to $E_n$, but the endpoints give a vanishing contribution because $E_n-
V(A_n)=E_n-V(B_n)=0$, where the potential $V(x)=x^2(ix)^\delta$.

Let us substitute (\ref{e30}) into (\ref{e5}) and examine the sum over $n$ of
the first term in the curly brackets in (\ref{e29}). Letting $r=\sqrt{E_n}/
\epsilon$, the sum over $n$ in (\ref{e5}) becomes the integral
\begin{eqnarray}
\frac{1}{\pi}\int_0^\infty dr\,\cos\left[r(x-y)\right]=\delta(x-y),
\label{e31}
\end{eqnarray}
which is precisely the statement of completeness. Of course, this completeness
result is only an approximation because the WKB approximation is only accurate
for large $n$. Thus, the early terms in the sum over $n$ introduce an error in
this result. This error is equivalent to changing the lower limit in the
integral in (\ref{e31}) from $0$ to some other finite number $L$. However, the
error that arises when $0$ is replaced by $L$ is finite, and when $x$ is near
$y$, this error is small and may be neglected in comparison with $\delta(x-y)$,
which is singular at $x=y$.

The second term in the curly brackets in (\ref{e29}) does not contribute to the
sum over $n$ in (\ref{e5}) because there is a factor of $(-1)^n$ in this term.
The rapid oscillation produced by this factor causes a cancellation between 
successive terms in the summation. As a result, this term contributes negligibly
to the sum and the delta function result in (\ref{e31}) remains unchanged.

Next, we turn to the evaluation of the $\mathcal{C}$ operator in (\ref{e6}).
Because there is no factor of $(-1)^n$ in this sum, it is now the {\it first}
term in curly brackets in (\ref{e29}) that is unimportant and it is the {\it
second} term that contributes. Converting this sum to an integral using
(\ref{e30}), we obtain in coordinate space
\begin{eqnarray}
\mathcal{C}(x,y)=\frac{1}{\pi}\int_0^\infty dr\,\cos\left[r(x+y)+iKr^\frac{4+
\delta}{2+\delta}\epsilon^\frac{2}{2+\delta}\right],
\label{e32}
\end{eqnarray}
where we have used the result that
\begin{eqnarray}
I_n-I_n^*=iKE_n^\frac{4+\delta}{4+2\delta}
\label{e33}
\end{eqnarray}
with the $n$-independent constant $K$ given by
\begin{eqnarray}
K=\frac{\sqrt{\pi}\Gamma\left(\frac{3+\delta}{2+\delta}\right)}{\Gamma\left(
\frac{8+3\delta}{4+2\delta}\right)}\sin\left(\frac{\pi\delta}{4+2\delta}\right).
\label{e34}
\end{eqnarray}

For small $\epsilon$ the cosine function  in (\ref{e32}) can be expanded into a
two-term Taylor series and the first integral can be done as in (\ref{e31}):
\begin{eqnarray}
\mathcal{C}(x,y)=\delta(x+y)-iK\epsilon^\frac{2}{2+\delta}\frac{1}{\pi}\int_0^
\infty dr\, r^\frac{4+\delta}{2+\delta}\sin\left[r(x+y)\right].
\label{e35}
\end{eqnarray}
To evaluate the remaining integral in (\ref{e35}), we transform to momentum
space:
\begin{eqnarray}
\mathcal{C}(p,q)=\int dx\,e^{-ipx}\int dy\,e^{iqy}\mathcal{C}(x,y).
\label{e36}
\end{eqnarray}
The Fourier transform of the first term in (\ref{e35}) is $2\pi\delta(p+q)$,
which is just the parity operator $\mathcal{P}(p,q)$ in momentum space. The
Fourier transform of the second term in (\ref{e35}) is
\begin{eqnarray}
&&-iK\epsilon^\frac{2}{2+\delta}\int dx\,e^{-ipx}\int dy\,e^{iqy}\int_0^\infty
\frac{dr}{\pi}r^\frac{4+\delta}{2+\delta}\frac{1}{2i}\left[e^{ir(x+y)}
-e^{-ir(x+y)}\right]\nonumber\\
&&\quad =-\frac{K}{2\pi}\epsilon^\frac{2}{2+\delta}\int dx\int dy\int_0^\infty
dr\,r^\frac{4+\delta}{2+\delta}\left[e^{i(r-p)x+i(q+r)y}-e^{-i(p+r)x+i(q-r)y}
\right]\nonumber\\
&&\quad=-2\pi K\epsilon^\frac{2}{2+\delta}\left[\theta(p)p^\frac{4+\delta}{2
+\delta}-\theta(-p)(-p)^\frac{4+\delta}{2+\delta}\right]\delta(p+q),
\label{e37}
\end{eqnarray}
where $\theta(t)=0\,(t<0)$, $\theta(t)=1\,(t\geq0)$ is the step function. Thus,
our result for the WKB approximation to the $\mathcal{C}$ operator in momentum
space is
\begin{eqnarray}
\mathcal{C}(p,q)=\mathcal{P}(p,q)-\epsilon^\frac{2}{2+\delta}\frac{\sqrt{\pi}
\Gamma\left(\frac{3+\delta}{2+\delta}\right)}{\Gamma\left(\frac{8+3\delta}{4+2
\delta}\right)}\sin\left(\frac{\pi\delta}{4+2\delta}\right)\left[\theta(p)
p^\frac{4+\delta}{2+\delta}-\theta(-p)(-p)^\frac{4+\delta}{2+\delta}\right]
\mathcal{P}(p,q).
\label{e38}
\end{eqnarray}

Finally, we rewrite this result in the exponential form $\mathcal{C}=e^Q\mathcal
{P}$ in (\ref{e13}), set $\epsilon=1$, and identify the function $Q$:
\begin{eqnarray}
Q(p,q)=-\frac{\sqrt{\pi}\Gamma\left(\frac{3+\delta}{2+\delta}\right)}{\Gamma
\left(\frac{8+3\delta}{4+2\delta}\right)}\sin\left(\frac{\pi\delta}{4+2\delta}
\right)\left[\theta(p)p^\frac{4+\delta}{2+\delta}-\theta(-p)(-p)^\frac{4+\delta}
{2+\delta}\right].
\label{e39}
\end{eqnarray}
This is the principal result of our investigation.

\section{Discussion and Conclusions}
\label{s3}

We have found a physical-optics WKB approximation to the $\mathcal{C}$ operator
in momentum space for the Hamiltonian in (\ref{e1}). Note that the results in
(\ref{e38}) and (\ref{e39}) are valid for all $\delta\geq0$. This represents a
major advance over previous work; previously, the $\mathcal{C}$ operator was
only determined for the special case of a cubic interaction term $\delta=1$.
The methods used earlier were perturbative and were not as powerful as the
nonperturbative semiclassical methods used in this paper.

The function $Q(p,q)$ exhibits the required symmetry properties mentioned
earlier, namely, that it be an even function of $x$ and an odd function of $p$.
Furthermore, because the sine function in (\ref{e38}) and (\ref{e39}) vanishes 
when $\delta=0$, we see that the $\mathcal{C}$ operator becomes identical with
the parity operator $\mathcal{P}$ when the Hamiltonian is Hermitian in the
conventional sense. The vanishing of $Q$ for the $\delta=0$ case can be traced
back to the fact that for the harmonic oscillator the integral $I$ in
(\ref{e26}) is real. Thus, $I-I^*=0$ in (\ref{e28}) and (\ref{e29}).

Note that the results in (\ref{e38}) and (\ref{e39}) simplify substantially when
$\delta$ is an odd integer because the step function $\theta$ drops out of these
formulas. For $\delta=2k+1$ we have
\begin{eqnarray}
Q(p,q)=-\frac{\sqrt{\pi}\Gamma\left(\frac{4+2k}{3+2k}\right)}{\Gamma\left(\frac{
11+6k}{6+4k}\right)}\sin\left(\frac{\pi(1+2k)}{6+4k}\right)p^\frac{5+2k}{3+2k}.
\label{e40}
\end{eqnarray}
When $k=0$ (the case of a cubic theory) we have
$$Q(p,q)=-\frac{\sqrt{\pi}\Gamma(1/3)}{5\Gamma(5/6)} p^{5/3},$$
and when $k=\infty$ we have the especially simple result that
$$Q(p,q)=-2p.$$
Another special case is $\delta=2$. For this case the Hamiltonian (\ref{e1}) is
quartic and
$$Q(p,q)=-\frac{\sqrt{2\pi}\Gamma(1/4)}{6\Gamma(3/4)}\left[\theta(p)p^{3/2}
-\theta(-p)(-p)^{3/2}\right].$$

We hope to generalize the advance reported in this paper to noncubic $\mathcal{P
T}$-symmetric quantum field theories, such as a $-g\varphi^4$ theory. A $-g
\varphi^4$ quantum field theory in four-dimensional space-time is a remarkable
model because it has a positive spectrum, is renormalizable, is asymptotically
free~\cite{BMS}, and has a nonzero one-point Green's function $G_1=\langle
\varphi\rangle$. If we can perform the corresponding semiclassical
approximation for a $-g\varphi^4$ quantum field theory, we may be able to
elucidate the dynamics of the scalar sector in a variant of the standard
electroweak model and possibly even predict the mass of the Higgs particle.

\begin{acknowledgments}
CMB is grateful to the Theoretical Physics Group at Imperial College for its
hospitality and he thanks the U.K. Engineering and Physical Sciences Research
Council, the John Simon Guggenheim Foundation, and the U.S.~Department of Energy
for financial support.
\end{acknowledgments}


\begin{enumerate}


\bibitem{BB} C.~M.~Bender and S.~Boettcher, Phys.~Rev.~Lett.
{\bf 80}, 5243 (1998).


\bibitem{BBM} C.~M.~Bender and S.~Boettcher and P.~N.~Meisinger,
J.~Math.~Phys. {\bf 40}, 2201 (1999).


\bibitem{R1} E. Caliceti, S. Graffi, and M. Maioli, Comm. Math. Phys. {\bf 75},
51 (1980).


\bibitem{R2} G.~L\'evai and M.~Znojil, J.~Phys. A{\bf 33}, 7165 (2000), and
references therein.


\bibitem{R3} B.~Bagchi and C.~Quesne, Phys.~Lett. A{\bf 300}, 18 (2002).


\bibitem{R4} Z.~Ahmed, Phys.~Lett. A{\bf 294}, 287 (2002);
G.~S.~Japaridze, J.~Phys.~A{\bf 35}, 1709 (2002);
A.~Mostafazadeh, J.~Math.~Phys. {\bf 43}, 205 (2002); {\em ibid}.
{\bf 43}, 2814 (2002); D.~T.~Trinh, PhD Thesis, University of Nice-Sophia
Antipolis (2002), and references therein.


\bibitem{DDT} P.~Dorey, C.~Dunning and R.~Tateo, J.~Phys.~A {\bf 34} L391
(2001); {\em ibid}. {\bf 34}, 5679 (2001).


\bibitem{AM} A. Mostafazadeh, J.~Math.~Phys.~{\bf 43}, 3944 (2002).


\bibitem{BBJ} C.~M.~Bender, D.~C.~Brody, and H.~F.~Jones,
Phys.~Rev.~Lett. {\bf 89}, 270402 (2002).


\bibitem{BBJ2} C.~M.~Bender, D.~C.~Brody, and H.~F.~Jones,
Am.~J.~Phys. {\bf 71}, 1095 (2003).


\bibitem{BMW} C.~M.~Bender, P.~N.~Meisinger, and Q.~Wang,
J.~Phys.~A {\bf 36}, 1973 (2003).


\bibitem{HH} C.~M.~Bender, G.~V.~Dunne, P.~N.~Meisinger, and M.~\d{S}im\d{s}ek,
Phys.~Lett.~A~{\bf 281}, 311-316 (2001).


\bibitem{BBRR} C.~M.~Bender, J.~Brod, A.~T.~Refig, and M.~E.~Reuter, J. Phys. A:
Math. Gen. (to be published).


\bibitem{AAAA} C.~M.~Bender, D.~C.~Brody, and H.~F.~Jones, hep-th/0402011.


\bibitem{BBBB} C.~M.~Bender, D.~C.~Brody, and H.~F.~Jones, hep-th/0402183, to
be published in Physical Review D.


\bibitem{BO} C.~M.~Bender and S.~A.~Orszag, {\it Advanced Mathematical Methods
for Scientists and Engineers}, (McGraw-Hill, New York, 1978), Chap.~10.


\bibitem{BMS} C.~M.~Bender, K.~A.~Milton, and V.~M.~Savage,
Phys.~Rev.~D~{\bf 62}, 85001 (2000).


\end{enumerate}
\end{document}